# Local structure and dynamics in colloidal fluids and gels

Takehiro Ohtsuka[1], C. Patrick Royall[1,2(a)] and Hajime Tanaka[1,2(b)]

[1] Institute of Industrial Science, University of Tokyo - Meguro-ku, Tokyo 153-8505, Japan

[2] School of Chemistry, University of Bristol - Bristol BS8 1TS, UK

**Abstract**

*Gels in soft-matter systems are an important nonergodic state of matter. We study a colloid-polymer mixture which is quenched by increasing the polymer concentration, from a fluid to a gel. Using confocal microscopy, we study both the static structure and dynamics in three dimensions (3D). Between the dynamically arrested gel and ergodic fluid comprised of isolated particles we find an intermediate "cluster fluid" state, where the "bonds" between the colloidal particles have a finite lifetime. The local dynamics are reminiscent of a fluid, while the local structure is almost identical to that of the gel. Simultaneous real-time local structural analysis and particle tracking in 3D at the single-particle level yields the following interesting information. Particles in the clusters move in a highly correlated manner, but, at the same time, exhibit significant dynamical heterogeneity, reflecting the enhanced mobility near the free surface. Deeper quenching eventually leads to a gel state where the "bond" lifetime exceeds that of the experiment, although the local structure is almost identical to that of the "cluster fluid".*

In addition to its relevance to many everyday applications the colloidal fluid-gel transition can serve as a model system for slow dynamics, due to the well-defined thermodynamic temperature of colloidal dispersions. Adding polymer to a colloidal suspension can induce effective attractions between the colloids. This depletion attraction, which drives the phase behaviour, results from the polymer entropy, since the polymer free volume is maximized when the colloids approach one another [1]. Controlling the state point by adding polymer in this way allows the system to be quenched from high effective temperatures, to dynamical arrest, which at moderate colloid concentrations results in a gel. At small polymer-colloid size ratios $q <0.25$, the addition of polymer leads to either phase separation to a gas-crystal phase coexistence [2–4], or to gelation [2,5], which may be interpreted as arrested phase separation [6–11]. Colloidal gels can also be viewed as a metastable state whose equilibrium phase is the fluid-crystal coexistence, with implications for generic systems with short-ranged attractions [10,12,13], such as globular proteins [3,14,15] and their crystallisation [16].

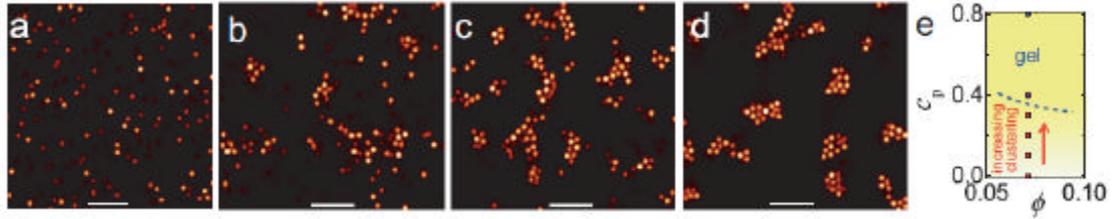

*Fig. 1: (Colour on-line) Confocal microscopy images at various polymer concentrations. (a) $c_P=0$, (b) $c_P=0.2$ gl$^{-1}$, (c) $c_P=0.3$ gl$^{-1}$ and (d) $c_P=0.8$ gl$^{-1}$. Bars = 10 µm. (e) State diagram. Shading and the dashed line are guides to the eye. Ergodic states are shown as red squares, gel as a blue circle.*

Recently, considerable interest has focussed on the role of electrostatic interactions in gelation [5,15,17], which may be important in the stabilisation of finite clusters [5,17,18]. Clusters composed of micron-sized colloidal particles [5,15,17], and relatively larger bead-like structures composed of proteins are readily visible using optical microscopy; however the interpretation of smaller clusters in protein solutions [14] has recently been called into question [19]. Here we consider a system which more closely resembles "sticky spheres" [12] since the electrostatic interactions appear very weak. Colloidal dispersions provide the level of structural and dynamic detail usually available only to computer simulations, since the particles can be structurally [20] and dynamically [21–24] resolved in 3D with confocal microscopy. We present results on structure-mobility correlations between near-neighbour particles in relatively dilute (ergodic) colloidal fluids and gels at the singleparticle level. We show that the local structure of the system is a strong function of polymer concentration in the ergodic fluid state, but that dynamic slowing at higher polymer concentration corresponds to little structural variation in quantities such as the radial distribution function. Recently, we underlined the importance of considering larger structures in colloidal gels at higher density (colloid volume fraction $f=0.35$) in which near-neighbour measures such as bond order parameters showed little structural change [25]. However, at the volume fraction considered here ($f=0.071\pm0.005$), near-neighbour bond order parameters provide a suitable means by which to characterise the local structure [17,26].

Dynamic slowing, on the other hand, occurs at higher polymer concentration (deeper quenching). At shallow quench depths, the system is an equilibrium fluid, with some evidence of clustering. While deep quenches lead to a gel with long-lived bonds, at intermediate polymer concentrations we find a state whose local structure resembles the gel but yet remains ergodic. We used poly(methyl methacrylate) (PMMA) colloids sterically stabilized with polyhydroxyl steric acid. The colloids were labelled with fluorescent rhodamine dye. To closely match the colloid density, such that very little sedimentation occurred on a time scale of days, we used a solvent mixture of cis-decalin and tetrachloro ethylene (TCE). We observed some swelling of the colloids, and believe that this is due to the uptake of TCE solvent, leading to a small refractive-index mismatch and van der Waals attraction between the colloids. This solvent has a low relative dielectric constant around 2.3. Solvents of similar dielectric constant exhibit weak charging of around 10 electronic charges per colloid [27]. Even these small charges might lead to significant electrostatic interactions, which can be determined using the radial distribution function g(r) [28].

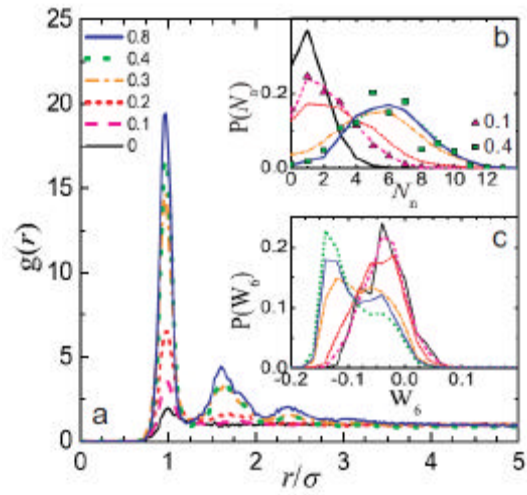

*Fig. 2: (Colour on-line) Local static structure. (a) Radial distribution functions. (b) Distribution of the coordination number $N_n$, $P(N_n)$. Lines denote large images, symbols small images from which dynamic data were taken. (c) Distribution of the bond order parameter $W_6$ (see text), $P(W_6)$. The legend denotes differing polymer concentrations in $gl^{-1}$, common to plots of (a)–(c).*

In particular, in this "energetic fluid" regime, the relation $g(r) \sim \exp(-\beta u(r))$, where $\beta = 1/k_BT$, where $k_BT$ is the thermal energy, and $u(r)$ is the pair interaction, is rather accurate [28,29]. In the case of electrostatic repulsions, we would therefore expect $g(r) < 1$ [30], which is not seen, fig. 2(a), black line. This suggests that any electrostatic repulsions are very weak. Instead, the van der Waals interactions noted above give rise to a small attraction. With Monte Carlo simulation, we estimate this attraction to be $(0.75 \pm 0.25)k_BT$ at contact, where $k_BT$ is the thermal energy [28]. Calculations based on the refractive-index mismatch are consistent with this value [31].

Following swelling, the colloids had a diameter s = 2.25±0.05 μm with around 3% polydispersity, as determined from static light scattering and analysis of radial distribution functions (RDF), fig. 2(a). All samples had a colloid volume fraction f = 0.071±0.005. The polymer used was polystyrene with a molecular weight of 8.6×10$^6$ and a polydispersity ratio $M_w/M_n$ =1.17, where $M_w$ and $M_n$ are the weight- and number-averaged molecular weight, respectively. The polymer underwent some swelling in these "good solvent" conditions. Although the Asakura-Oosawa (AO) model assumes ideal interactions between the polymers [1], for small polymer-colloid size ratios q [32] the AO model can still provide good agreement with experimental data, if an "effective" polymer size is taken [28]. Following an identical procedure to that in [28], we estimate an effective polymer radius of gyration $R_G$ ~ 125 nm, which we take as bond length [33], yielding a polymer-colloid size ratio q=0.11. We used a Leica SP5 confocal microscope, fitted with a resonant scanner, and tracked the coordinates of each particle with a precision of around 100nm [20,30].

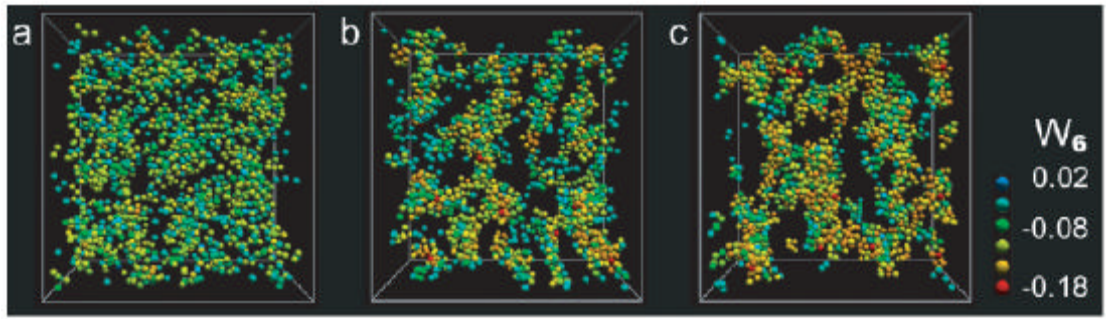

*Fig. 3: (Colour on-line) Rendered 3D coordinates for (a) $c_p$ =0.2 gl$^{-1}$, (b) $c_p$ =0.3 gl$^{-1}$, and (c) $c_p$ =0.8 gl$^{-1}$. The colours denote $W_6$ values as shown in the legend. All particles show 0.8 actual size.*

We begin with confocal microscopy images of the system, as shown in fig. 1. Under the weak van der Waals attraction, slight clustering is observed in the system without added polymer (fig. 1(a)). The addition of even a relatively small amount of polymer ($c_p$ =0.2gl$^{-1}$) drives significant clustering (fig. 1(b)), while at a polymer concentration $c_p$ =0.3gl$^{-1}$ relatively few free particles are seen (fig. 1(c)). At $c_p$ =0.8gl$^{-1}$ the system is a gel (fig. 1(d)) with essentially no free particles. Next we consider the structure. Figure 2(a) shows colloid-colloid RDFs calculated from particle coordinates, from a volume of 50×50×50 μm. Upon quenching, these show a considerable increase in the first peak, due to short-ranged attractions induced by the polymer. Higher-order maxima correspond to the formation of clusters. In the colloidal fluid at low polymer concentrations ($c_p$ _ 0.3gl$^{-1}$), the height of the first peak is a super-linear function of the polymer concentration, characteristic of the "ergodic fluid" regime [28,29]. At higher polymer concentrations, we see comparatively little development (fig. 3) for $c_p$ _0.4gl$^{-1}$.

Our next consideration is to impose a bond length. The polymer presents a natural length scale for the bond length, since the interaction range in the Asakura-Oosawa model is 2$R_G$ [1]. Figure 1(b) shows the distribution of the number of nearest neighbours $N_n$, which we denote $P(N_n)$. Like the g(r) data, in the energetic fluid, $P(N_n)$ shows a strong response to the polymer concentration for $c_p \gtrsim 0.3$ gl$^{-1}$ after which there is little change. Identifying bonds also allows us to obtain bond order parameters. In particular, the invariant $W_6$ takes negative values in the case of local fivefold symmetry, and is close to zero for local crystalline environments [34]. $W_6$ is calculated for each particle and the distribution of $W_6$, $P(W_6)$, is presented in fig. 1(c) [35]. There is a significant negative shift in $P(W_6)$ between $c_p =0.2$ and $0.4$ gl$^{-1}$. Although there is little change in the essential features of g(r) and coordination number for $c_p \gtrsim 0.4$ gl$^{-1}$, $P(W_6)$ undergoes some changes between these state points, suggesting a degree of local re-ordering. The connection between $W_6$ and structure is revealed in fig. 3, where the particles are colour-coded according to $W_6$. This 3D rendering is suggestive of some overall connectivity in fig. 3(c) ($c_p =0.8$ gl$^{-1}$); however it is hard to draw definite conclusions from a $50\times50\times50$ μm$^3$ sample volume.

The shift in $W_6$ distribution around $c_p =0.3$–$0.8$ gl$^{-1}$ appears to be correlated with the development of large clusters. In other words, it seems that the particles located towards the middle of the "clusters" or "branches" have a more negative $W_6$, which we interpret as a stronger degree of fivefold symmetry [34]. Furthermore, those particles in the middle of the "clusters" or "branches" have more neighbours than those on the surface. Recalling the short-ranged nature of the attraction (q ~ 0.11), the coordination number provides an estimate of the potential energy of each particle. Thus, it is tempting to correlate a relatively high degree of fivefold symmetry with low local potential energy. Since the equilibrium state is expected to be the fluid-crystal coexistence [4], we draw a parallel with those glasses whose ground state is a crystal, the formation of which may be frustrated by local fivefold symmetry [36,37]. We noted in fig. 1(b) that some weak clustering like behaviour is apparent. While these loosely bound structures are distinct from compact, well-defined clusters, we nevertheless note that, under our definition of a bond length, large groups of particles appear transiently bound together in the fluid state $c_p \gtrsim 0.3$ gl$^{-1}$. We plot the radius of gyration $r_G$ as a function of the number of bound particles N in fig. 4(a). Although our local analysis is far from the thermodynamic limit, we estimate a local fractal dimension $d_f$ from the slope in fig. 4(a), in a similar spirit to [38]. The result is shown in 4(b). Without added polymer, the local fractal dimension is about $d_f =1.8$, which despite the finite cluster lifetime and small cluster size is close the value for diffusion-limited cluster aggregation (DLCA) [39]. At higher polymer densities, $d_f$ initially increases, demonstrating a denser packing. Above $c_p \gtrsim 0.4$, no further increase is seen in $d_f$, which may be associated with gelation: the development of local fivefold symmetric structures (see above) may inhibit crystallisation, which would lead to densification and further increase of $d_f$.

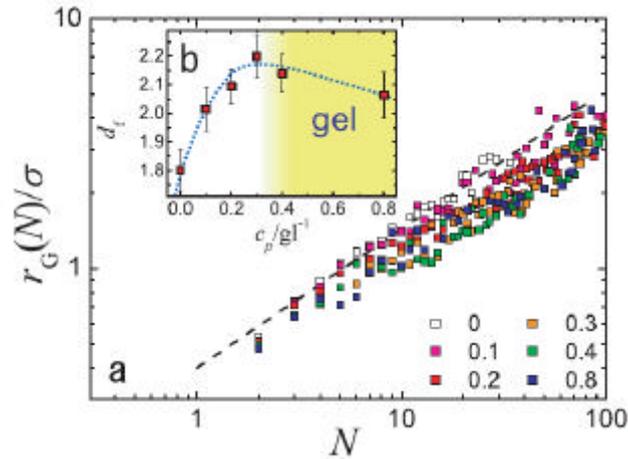

*Fig. 4: Local fractal dimension of connected regions. (a) Radius of gyration as a function of size. The dashed line corresponds to the DLCA fractal dimension (1.8). (b) Local fractal dimension $d_f$ as a function of polymer concentration. The dotted line is a guide for the eye.*

We have noted that for $c_p =0.8$ gl$^{-1}$, our dynamical measurements provide strong evidence of gelation. Now all our structural probes reveal essentially no difference between $c_p =0.4$ and $c_p =0.8$ gl$^{-1}$; moreover, the fact that around $c_p \sim 0.4$ gl$^{-1}$, the structure ceases to respond to polymer addition suggests that some change occurs.

Furthermore, the long-time diffusion constant decreases strongly for $c_p \sim 0.4$ gl$^{-1}$, which leads us to interpret $c_p =0.4$ gl$^{-1}$, as a gel. Now we turn to the local dynamics, which are determined using time-resolved 3D coordinate tracking. In order to track the diffusing particles in time as well as space, it was necessary to considerably reduce the volume sampled to 20×20×10 μm, which we imaged every dt=0.73 s. This is fast enough to follow the dynamics at the single-particle level, since the characteristic free diffusion time over a radius is about 13 s. On the other hand, clearly, it is important to be sure that such a small sample of N~50 particles is indeed representative of the bulk. In fig. 2(b) we compare the distribution of the coordination number (symbols denote small images) for two state points ($c_p =0.1$ and $0.4$ gl$^{-1}$), which provides a sensitive measure since fewer neighbours may be found in the case of smaller images. There is very little change in the distribution for both states which gives us confidence that our analysis is reasonably robust.

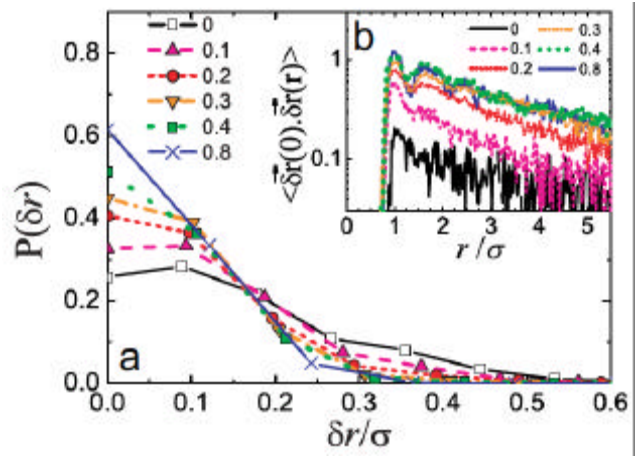

*Fig. 5: (Colour on-line) (a) Distribution of the displacement dr. (b) Spatial correlation in displacement vectors <dr(r)>.<dr(0)>*

Identifying particles between successive images, we determined displacements between frames (dt) in a similar way to [21], although we treat displacements in x, y, z on the same footing. The accuracy of our coordinate tracking leads to a lower bound of the measured displacement approaching 0.1s. Since our measurements are in the diffusive regime, we define dr as the displacement scaled by the square root of the relative viscosity $\eta_r = \eta/\eta_0$, where $\eta$ and $\eta_0$ are the viscosity of the polymer solution and the solvent, respectively. Figure 5(a) is a distribution of dr (=|dr|). As may be expected, by increasing the polymer concentration, the mobility falls. In the case in which $c_P = 0.8$ gl$^{-1}$, the displacement is quite close to our resolution, suggesting relatively little movement, as expected for a strongly quenched gel.

We see no evidence of bond breakage on the experimental time scale. In fig. 5(b) we plot the spatial correlation of particle motion which extends well beyond nearest neighbours and becomes stronger with $c_P$. The oscillatory feature observed for $c_P \geq 0.3$ gl$^{-1}$ corresponds to the structural correlations of g(r) (fig. 1(a)). This spatial correlation also tends to collapse onto the same line for $c_P \geq 0.3$ gl$^{-1}$, much like g(r).

We can also see the correlation in displacement direction between individual particles and their neighbours by using $\langle dr_i \cdot dr_n \rangle / (|dr_i||dr_n|) = \cos\theta_i$, where $dr_i$ is the displacement of particle i per dt, $dr_n$ is the average displacement of its neighbours, and $\theta_i$ is the angle between $dr_i$ and $dr_n$ (see fig. 6(b), inset). Figure 6(a) plots the average of $\cos\theta_i$ over all particles of coordination number $N_n$, $\cos\theta_i$, as a function of $N_n$. These results suggest a strong link between the number of neighbours and the cooperativity in particle motion. In other words, the relative motion of particles becomes increasingly cooperative with more neighbours at all polymer concentrations. The fact that the local structural development accompanies a strong enhancement of the spatial cooperativity of particle motion indicates a strong structure-dynamics correlation: In the way that for $c_p$ =0.3, 0.4 and 0.8$gl^{-1}$, the RDFs almost collapse upon one another (fig. 2(a)), so too do the cos $\theta_i(N_n)$ in fig. 6(a). In other words, tightly bound regions behave more or less as a rigid body. Note that in the case of the deeply quenched gel, $c_p$ =0.8$gl^{-1}$, strong correlations are already apparent even in the case of $N_n$ =1, i.e., in this case the bonds are quasi-permanent.

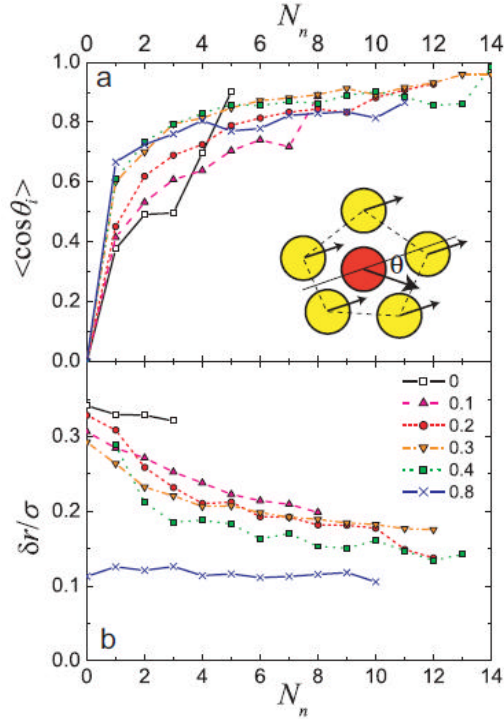

*Fig. 6: (a) Correlation between cos $\theta_i$ and coordination number $N_n$. The inset in (a) shows $\theta$, the angle between the displacement of a central particle (red) and its neighbours. (b) dr as a function of the coordination number $N_n$. The legend denotes differing polymer concentrations in $gl^{-1}$.*

The rendering in fig. 3 suggests that the coordination number $N_n$ provides a way to differentiate particles on the "surface" and those towards the middle of the "branches". We now plot the displacements dr as a function of $N_n$ in fig. 6(b). At all state points, except $c_p$ =0.8$gl^{-1}$, we see a correlation between displacement and coordination number. Low coordination numbers correspond to particles on the "surface"; these particles are more mobile than

Colours and types of lines are common between both plots. those towards the middle of the "branches". This enhanced mobility near the free surface is quite natural, and also observed in thin films of glass-forming materials [41]. This link to the enhanced mobility on a free surface of glass is also compatible with a scenario that gel is formed as a result of dynamic arrest by vitrification during phase separation [6–9,11]. Furthermore, the identification of the role of the surface in dynamical heterogeneity in colloidal gels is consistent with computer simulations [42]. The absence of such a dynamic heterogeneity in the lowest temperature gel ($c_p$ =0.8gl$^{-1}$) is indicative of the long bond lifetime [43,44]. The entire structure may be thought of as "frozen".

Before closing, some comments on the longer-term dynamics are in order. We have focussed on the short term mobility, and, as figs. 5(a) and 6(b) clearly show, all state points other than $c_p$ =0.8gl$^{-1}$ exhibit considerable local mobility. However, even the final-state point, $c_p$ =0.8gl$^{-1}$, which is definitely a gel, exhibits some long-wavelength motion (network fluctuations). Thus, in these macroscopic dilute experimental systems, a quantity such as the mean squared displacement may not necessarily result in a plateau. This forms a marked difference between gels and glasses.

We have investigated the structure and dynamics of a colloidal fluid, with short-range attractions, quenched to gelation, at the single-particle level. The most dramatic structural development occurs in the energetic fluid regime, $c_p$ _0.3gl$^{-1}$. The radial distribution function, coordination number and bond order parameter $W_6$ are strong functions of polymer concentration in the ergodic fluid, but show little change for $c_p$ > 0.3gl$^{-1}$. Conversely, the absolute mobility shows comparatively little change until further quenching to $c_p$ =0.8gl$^{-1}$, where the bond lifetime exceeds the experimental time scale. Particle level dynamics show a strong dependence upon the local environment. As the system approaches gelation, locally dense regions are formed. In the interior of these, with a high coordination number, mobility is suppressed, compared to particles on the "surface". This enhanced mobility on surface may play a significant role in the restructuring and the aging of gels.

We now consider possible scenarios for the behaviour we have observed. Gelation may be connected to an underlying gas-liquid spinodal [6–11]. While the deeply quenched gel at $c_p$ =0.8gl$^{-1}$ shows a bicontinuous structure consistent with spinodal decomposition, we are inclined to interpret our results more specifically as viscoelastic spinodal phase separation [8]. For the cluster fluid, on the other hand, there are two possibilities: 1) arrested nucleation or droplet spinodal decomposition. Such behaviours are observed in protein solutions [5,14,17,45,46] and polymer solutions (moving droplet state) [47]. These examples considered well-defined clusters of up to tens of thousands of particles [45], or even more in the case of the protein beads [3,46] and polymer droplets [47], orders of magnitude greater than the weak clustering observed here. The formation of these large droplets is a consequence of arrested phase separation [48]. Thus, even if the clusters do not grow, (arrested) phase separation may still be the underlying mechanism. 2) Clustering due to the competition between relatively weak attractions and entropy. In this case, the cluster fluids should be in an equilibrium stable state. The present results alone do not provide a definitive answer to this fundamental question, but it is tempting to take scenario 2), since the clusters are internally mobile and not dynamically arrested (see movies). While future work will identify the origin of clustering in this system, we note the possible relevance of weak electrostatic interactions in stabilising the weak cluster-like structures we observe [18].

Finally we mention a mechanism preventing crystallization. Although the underlying state is expected to be the fluid-crystal coexistence [4], we find no evidence of crystallisation, rather we interpret the negative shift of the bond order parameter $W_6$ as an increase in fivefold symmetry with polymer concentration, which may impede crystallisation [36,37]. A link between this suppression of crystallisation and particles acquiring a relatively large coordination number, in other words being in a local potential energy well, was recently emphasised [25], along with explicit local energy minima as a more sensitive alternative to $W_6$.

The authors are grateful to D. Derks for a kind gift of colloidal particles, and to W. Kob, M. Miller and T. Nikolai for helpful discussions. CPR acknowledges the Royal Society for financial support. HT acknowledges a grant-in-aid from the Ministry of Education Culture, Sports, Science and Technology, Japan.